\documentclass[12pt]{iopart}

\usepackage{graphicx}
\begin{document}

\title{Magnetoelectric effect due to local noncentrosymmetry}

\author{V P Sakhnenko and N V Ter-Oganessian}
\address{Institute of Physics, Southern Federal University,
194 Stachki Prospekt, Rostov-on-Don, 344090 Russia}
\ead{nikita.teroganessian@gmail.com}

\begin{abstract}
Magnetoelectrics often possess ions located in noncentrosymmetric
surroundings. Based on this fact we suggest a microscopic model of
magnetoelectric interaction and show that the spin-orbit coupling
leads to spin-dependent electric dipole moments of the electron
orbitals of these ions, which results in non-vanishing
polarization for certain spin configurations. The approach
accounts for the macroscopic symmetry of the unit cell and is
valid both for commensurate and complex incommensurate magnetic
structures. The model is illustrated by the examples of MnWO$_4$,
MnPS$_3$ and LiNiPO$_4$. Application to other magnetoelectrics is
discussed.
\end{abstract}


\maketitle

\section{Introduction}

Pierre Curie~\cite{Curie} was the first to predict the interplay
between magnetism and ferroelectricity. The macroscopic symmetry
consideration of the magnetoelectric (ME) effect was given by
Dzyaloshinskii only in 1959~\cite{Dzyaloshinskii}, whereas
experimentally it was discovered by Astrov in Cr$_2$O$_3$ in
1960~\cite{Astrov}. In the last decade whole new classes of
magnetoelectric materials were discovered and experimentally
studied (for a recent review see, for example,
\cite{ReviewSpiralTokura}).

Despite a long history of ME effect its microscopic origin is
still a subject of debate. The fact that ferroelectricity commonly
occurs in incommensurately modulated magnetically ordered phases
predominates the assumption and consideration of various complex
types of magnetic order such as screw, cycloidal, helix and others
in most of the microscopic models of
magnetoelectricity~\cite{ReviewSpiralTokura,ArimaSpinDrivenFerro}.
From the macroscopic crystal symmetry point of view the close
connection between the appearance of modulated and ferroelectric
phases in magnetoelectrics was recently pointed
out~\cite{SakhnenkoImproperFerroelectric}. Whereas electric
polarization indeed often occurs concurrently with complex
modulated spin structures it can also be induced by commensurate
magnetic order (such is the case, for example, in some rare-earth
manganates RMn$_2$O$_5$~\cite{Kimura125manganites}).

Currently two models of ME coupling are widely accepted in
literature. In the model by Sergienko et al.~\cite{SergienkoDMI}
electric polarization $\vec{P}\sim\ [\vec{S}_i\times
\vec{S}_{i+1}]$ is induced by the Dzyaloshinskii-Moriya (DM)
interaction between two magnetic ions $i$ and $i+1$ with
superexchange. In the spin current model by Katsura et al.~\cite{KatsuraSpinCurrent} the electric polarization is due to the
spin supercurrent $\vec{P}\sim\vec{e}_{ij} \times [\vec{S}_i
\times\vec{S}_{j}]$ with $\vec{e}_{ij}$ being the unit vector
connecting the sites $i$ and $j$. Interpretation of the
experimentally observed polarization using these models meets
difficulties though and was a subject of
critique~\cite{MochizukiSpinCurrent,MoskvinPRB}. It was
argued~\cite{MoskvinPRB} that the DM coupling~\cite{SergienkoDMI}
is 2 orders of magnitude weaker than what is needed to explain
experimental situation in magnetoelectric manganites. On the other
hand in the spin current model~\cite{KatsuraSpinCurrent} the
authors also overestimated up to 2 orders of magnitude the value
of the magnetically induced polarization~\cite{MoskvinPRB}.
Several other mechanisms of the ME effect were
proposed~\cite{MoskvinPRB,MoskvinEPJB} but as argued by the
authors themselves they are insufficient to explain
magnetoelectricity in some Cu$^{2+}$ magnetoelectrics.

From our point of view most of the microscopic models proposed so
far do not take into account the macroscopic crystal symmetry
mainly focusing on three-site clusters (two metal ions and oxygen)
and considering various spiral magnetic structures. The authors of
the DM model of ME effect~\cite{SergienkoDMI}, for example, apply
it to the case of rare-earth manganites RMnO$_3$, whereas it can
be shown that the macroscopic polarization vanishes when one
considers the orthorhombic symmetry of the unit
cell. Indeed, the authors take TbMnO$_3$ as an example using the
spiral magnetic structure available from neutron diffraction data.
They consider the polarization induced in the $x$ and $y$ chains
of Mn-O-Mn, but erroneously take into account only the $z=0$
plane. Indeed, when one considers also the $z=1/2$ plane (obtained
from the first one by $\sigma_z(00\frac{1}{2})$) the macroscopic
polarization cancels out.

The aim of any ME model is to find the mechanism of inversion
symmetry breaking by magnetic order. At the same time some authors
explicitly start with a centrosymmetric cluster such as, for
example, in the spin
current~\cite{KatsuraSpinCurrent,JiaOnodaNagaosaMOMcluster} or the
MeO$_n$ cluster model~\cite{MoskvinPRB}. This dismisses the fact
that in many magnetoelectrics some types of ions are located in
noncentrosymmetric surroundings already in the paramagnetic phase.
In this work using MnWO$_4$ as an example we suggest a microscopic
magnetoelectric coupling model taking into account macroscopic
symmetry of the unit cell and noncentrosymmetric surrounding of
the Mn$^{2+}$ ions. We then apply our microscopic model to
estimate the linear magnetoelectric coefficients in MnPS$_3$ and
LiNiPO$_4$ and discuss its application to other magnetoelectrics.

\section{Microscopic model}

\begin{figure}
\begin{indented}
\item[]\includegraphics{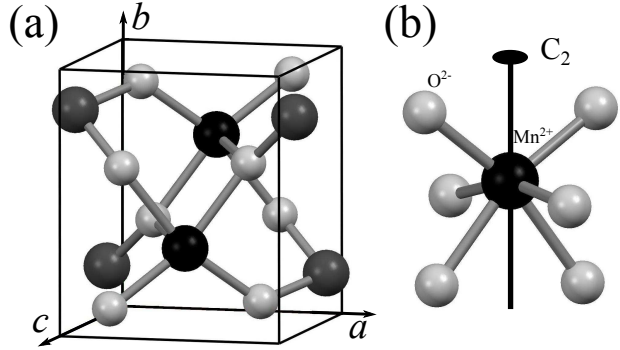}
\end{indented}
\caption{\label{fig:MWOStructure_and_Cluster} (a) The monoclinic unit cell of MnWO$_4$. Mn$^{2+}$, W$^{6+}$ and
O$^{2-}$ ions are shown by black, dark gray and light gray circles, respectively. (b) The Mn$^{2+}$ cluster
showing distorted oxygen octahedron. The rotational C$_2$ symmetry axis is indicated.}
\end{figure}

Wolframite MnWO$_4$ possesses a monoclinic structure at room
temperature (figure~\ref{fig:MWOStructure_and_Cluster}a) described by the space group $P2/c$ (C$_{2h}^4$). On
lowering the temperature it undergoes a sequence of magnetic phase
transitions at 13.5~K (T$_N$), 12.7~K (T$_2$) and 7.6~K (T$_1$),
leading to the appearance of magnetically ordered states
\textbf{AF3}, \textbf{AF2} and
\textbf{AF1}~\cite{Lautenschlaeger}, respectively. The low
temperature phase \textbf{AF1} is characterized by the wave vector
$\vec{k}=(1/4;1/2;1/2)$, whereas the incommensurate phases
\textbf{AF2} and \textbf{AF3} by $(-0.214;1/2;0.457)$. Electric polarization in MnWO$_4$ appears in the \textbf{AF2}
phase along the crystal $b$ axis~\cite{Taniguchi06}. The
phenomenological model of phase transitions in wolframite was
suggested earlier based on the assumption that the magnetic order
is driven by the instability in the $(1/4;1/2;1/2)$ point of the
Brillouin zone~\cite{SakhnenkoMWO}. However, for our purpose of
building a microscopic model of the ME effect we first start with
a hypothetical magnetic order with $\vec{k}=0$ in MnWO$_4$ (i. e.
without multiplication of the unit cell) and then consider the
real magnetic structure.

MnWO$_4$ contains two magnetic Mn$^{2+}$ ions Mn$_1$ and Mn$_2$ in
the unit cell located at positions $(0.5;0.6853;0.25)$ and
$(0.5;0.3147;0.75)$, respectively. In the following we define the
orthogonal $x$, $y$ and $z$ axes parallel to the $a$ axis,
parallel to the $b$ axis and perpendicular to both the $a$ and $b$
axes of the monoclinic cell, respectively. For the description of
the magnetic order with $\vec{k}=0$ we can introduce ferromagnetic
and antiferromagnetic order parameters
$\vec{F}=\vec{S}_1+\vec{S}_2$ and $\vec{A}=\vec{S}_1-\vec{S}_2$,
respectively, where $\vec{S}_1$ and $\vec{S}_2$ are the magnetic
moments of Mn$_1$ and Mn$_2$. Table~\ref{tab:GM-IRs} summarizes
the irreducible representations (IR) possessed by the $P2/c$ space
group in the center of the Brillouin zone and shows how the
components of $\vec{F}$, $\vec{A}$ and electric polarization
$\vec{P}$ transform according to the symmetry operations of the
group.
\begin{table}
\caption{\label{tab:GM-IRs}%
IR's of the $P2/c$ space group corresponding to $\vec{k}=0$. The last column lists the components
of $\vec{F}$, $\vec{A}$ and $\vec{P}$ according to the IR's upon
which they transform. Note that $\vec{F}$ and $\vec{A}$ are odd under time inversion.}
\begin{indented}
\item[]\begin{tabular}{@{}llll}
\br
IR & $C_{2y}\left(00\frac{1}{2}\right)$ & $I\left(000\right)$ & Order parameters\\
\mr
GM$^{1+}$ & ~1 & ~1 & $F_y$\\
GM$^{1-}$ & ~1 & -1 & $P_y$, $A_y$\\
GM$^{2+}$ & -1 & ~1 & $F_x$, $F_z$\\
GM$^{2-}$ & -1 & -1 & $P_x$, $P_z$, $A_x$, $A_z$\\
\br
\end{tabular}
\end{indented}
\end{table}
As evident from the table the following magnetoelectric
interactions are allowed by the macroscopic symmetry
\begin{equation}
\eqalign{ P_\mu A_xF_y,\qquad P_\mu A_yF_x,\qquad P_\mu A_yF_z,\qquad P_\mu A_zF_y,\cr
P_yA_\alpha F_\alpha,\qquad P_yA_xF_z,\qquad P_yA_zF_x,}\label{EQs:MagnetoelectricInteractions}
\end{equation}
where $\mu=x,z$ and $\alpha=x,y,z$. Thus any combination of
ferromagnetic and antiferromagnetic ordering produces electric
polarization. The polarization also arises in case when only one
of the Mn$^{2+}$ moments orders (e. g. when $\vec{S}_1\neq0$ and
$\vec{S}_2=0$, which corresponds to $\vec{F}=\vec{A}$). Such
collinear commensurate magnetic ordering gives no polarization
according to any of the microscopic models proposed so far, since
they are usually concentrated on noncollinear modulated magnetic
structures. Nevertheless, from the crystal symmetry point of view
such magnetic order breaks the crystallographic equivalency of
Mn$_1$ and Mn$_2$ atoms, which are connected by inversion (i.~e.
breaks inversion symmetry) and electric polarization arises.

\begin{figure}
\begin{indented}
\item[]\includegraphics{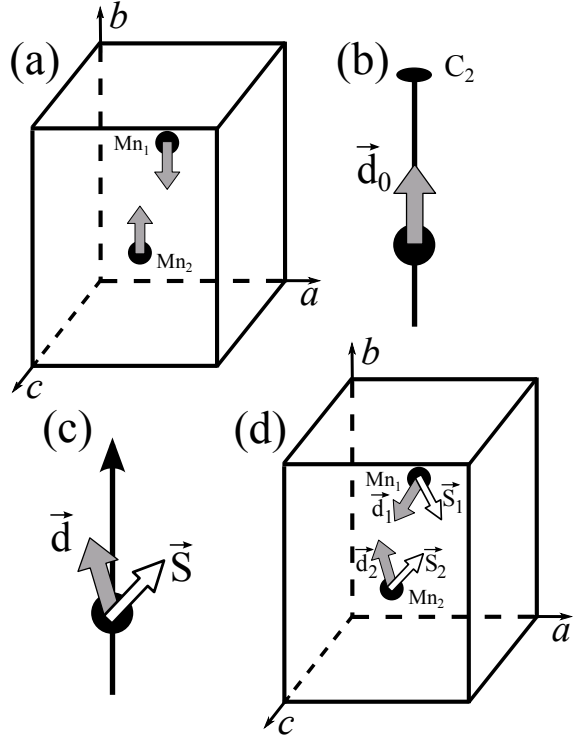}
\end{indented}
\caption{\label{fig:BoxIon} (a) Two Mn$^{2+}$ ions in the
monoclinic unit cell with opposing electric dipole moments
(gray arrows). (b) An ion (black circle) in C$_2$ crystal field
(indicated) possessing dipole moment $\vec{d}_0$ (gray arrow). (c) Same as (b)
but taking the spin degree of freedom into account. A spin
$\vec{S}$ (white arrow) deviating from the C$_2$ axis breaks the rotational
symmetry and changes the electric dipole moment to $\vec{d}$
deflecting it from the initial symmetry axis. (d) Two Mn$^{2+}$ ions in the
monoclinic unit cell with spin dependent electric dipole moments
$\vec{d}_1$ and $\vec{d}_2$ resulting in nonzero total dipole moment of the unit cell $\vec{d}_1+\vec{d}_2$.}
\end{figure}

In order to build a valid microscopic model of ME interaction we
proceed with the following consideration. First we note that the
Mn$^{2+}$ ions in MnWO$_4$ are located in noncentrosymmetric
positions. This can be seen from the fact that $I\left(000\right)$
interchanges Mn$_1$ and Mn$_2$ or by direct examination of oxygen
positions around Mn$^{2+}$ ions~\cite{Lautenschlaeger}. Since
$C_{2y}\left(00\frac{1}{2}\right)$ transforms each Mn$^{2+}$ ion
into itself their local crystal symmetry is C$_2$ as depicted in figure~\ref{fig:MWOStructure_and_Cluster}b. Thus the
manganese ions are located in polar surroundings and all their
electron's orbitals have electric dipole moment along the $y$
axis. Every monoclinic cell has two Mn$^{2+}$ ions with opposite
dipole moments, thus, preserving zero dipole moment of the unit
cell as schematically shown in figure~\ref{fig:BoxIon}a.

Figure~\ref{fig:BoxIon}b shows an ion in polar crystal environment
(local symmetry C$_2$). All electron orbitals possess, therefore,
an electric dipole moment. If one takes the spin degree of freedom
into account then the spin-orbit coupling alters the electric
dipole moment. When the spin directs away from the two-fold
rotational axis it breaks the rotational symmetry resulting in
modification of the dipole moment as schematically shown in
figure~\ref{fig:BoxIon}c. We now proceed from this simple physics
consideration to semiquantitative quantum mechanical treatment.

The manganese ions in MnWO$_4$ are located in distorted oxygen
octahedra as shown in figure~\ref{fig:MWOStructure_and_Cluster}b. Therefore we start with a Mn$^{2+}$ ion in octahedral
crystal field and for simplicity consider only the $e_g$ orbitals.
Inclusion of $t_{2g}$ levels is straightforward. Thus as a zeroth
order perturbation we have
\[
H_0|d_\gamma\rangle=E_d|d_\gamma\rangle,
\]
where $\gamma=z^2$ or $x^2-y^2$, $H_0$ is the hamiltonian
including octahedral splitting field and $E_d$ is the $e_g$ energy
level. Next, we include as perturbation the monoclinic crystal
field of C$_2$ symmetry assuming the polar axis along the $z$ axis
\begin{equation}\label{EQ:VcrystalField}
V_{CF}=c_zz+c_{xy}xy+c'_{xy}(x^2-y^2)+c'_{z}z^2
\end{equation}
and spin-orbit coupling
\begin{equation}\label{EQ:SOcoupling}
V_{SO}=-\lambda(\vec{L}\cdot \vec{S}).
\end{equation}
Here $c_z$, $c_{xy}$, $c'_{xy}$ and $c'_{z}$ are coefficients,
$\vec{L}$ is the angular momentum operator, $\vec{S}$ is the spin
and $\lambda$ is the spin-orbit coupling constant. Thus, the
perturbed hamiltonian has the form $H=H_0+V$, with
$V=V_{CF}+V_{SO}$.

The perturbation $V$ mixes the unperturbed $3d$ $e_g$ states with
other states and for simplicity it is sufficient to consider only
the $4p$ states $H_0|p_\alpha\rangle=E_p|p_\alpha\rangle$ with the
energy $E_p$, $\alpha=x,y,z$. The two-fold degeneracy is removed
in the first order and one can write for one of the perturbed
eigenvectors
\begin{equation}\label{EQ:PSIwavefunction}
|\psi\rangle=|0\rangle+\sum_\alpha A_\alpha |p_\alpha\rangle,
\end{equation}
where $A_\alpha$ are coefficients and $|0\rangle$ is an
eigenvector from the subspace spanned by $|d_\gamma\rangle$. The
electric dipole moment is given then by
\[
\vec{d}=\langle\psi |e \vec{r}|\psi \rangle=\sum_\alpha A_\alpha
\langle 0|e\vec{r}|p_\alpha\rangle + c.c.
\]
Assuming $|0\rangle=q_1 |d_{z^2}\rangle + q_2 |d_{x^2-y^2}\rangle$
we obtain in the first order of perturbation the $z$-component of
the electric dipole moment induced by the local crystal field
\[
d_{0z}=\frac{2q_1^2 c_ze   t_{z,z^2}^2}{\Delta},
\]
where $\Delta=E_d-E_p$ and $t_{\alpha,\gamma}=\langle p_\alpha |
\alpha |d_{\gamma}\rangle$. Due to local C$_2$ crystal field
symmetry $d_{0x}=d_{0y}=0$. Performing the perturbation up to the
third order we get the spin-independent first order contribution
and the spin-dependent part of the dipole moment as
\begin{equation}\label{EQs:DipoleMoment}
\eqalign{
d_{x}=-Q_{x}\left(\frac{\lambda}{\Delta}\right)^2S_xS_z, \cr
d_{y}=-Q_{y}\left(\frac{\lambda}{\Delta}\right)^2S_yS_z, \cr
d_{z}=d_{0z}+d_{0z}\left(\frac{\lambda}{\Delta}\right)^2(S_x^2+S_y^2),}
\end{equation}
where
\[
Q_\alpha=\frac{2q_1  c_ze t_{z,z^2}(q_1 t_{\alpha,z^2}+ q_2
t_{\alpha,x^2-y^2})}{\Delta}.
\]
Thus, in addition to the crystal field induced electric dipole
moment the spin-orbit coupling gives rise to spin-dependent
contribution.

We now use the obtained results for the above case of magnetic
ordering in MnWO$_4$ with $\vec{k}=0$. Each unit cell has two
Mn$^{2+}$ ions in opposite polar surroundings with polar axes
along $y$. Therefore, using~(\ref{EQs:DipoleMoment}) for both of them and performing the proper cyclic permutation of indices $x$, $y$ and $z$ we get the electric polarization
\begin{equation}\label{EQs:Polarizationk=0}
\eqalign{
P_{x}=-Q_{x}\frac{1}{2v}\left(\frac{\lambda}{\Delta}\right)^2(A_yF_x+A_xF_y), \cr
P_{y}=d_{0y}\frac{1}{v}\left(\frac{\lambda}{\Delta}\right)^2(A_xF_x+A_zF_z), \cr
P_{z}=-Q_{z}\frac{1}{2v}\left(\frac{\lambda}{\Delta}\right)^2(A_yF_z+A_zF_y),}
\end{equation}
where $v$ is the unit cell volume. Therefore, we obtain the same
polarization as the one implied by the ME
interactions~(\ref{EQs:MagnetoelectricInteractions}). Figure~\ref{fig:BoxIon}d schematically illustrates relations~(\ref{EQs:Polarizationk=0}) showing Mn$^{2+}$ ions with
spins $\vec{S}_1$ and $\vec{S}_2$ directed in such a way that the spin-dependent electric dipole moments
$\vec{d}_1$ and $\vec{d}_2$ result in nonzero total dipole moment of the unit cell $\vec{d}_1+\vec{d}_2$.

Other ME interactions present in~(\ref{EQs:MagnetoelectricInteractions}) and absent in~(\ref{EQs:Polarizationk=0}) can be obtained by performing the quantum perturbations to higher orders. It has
to be noted, that when deriving the spin-dependent dipole
moments~(\ref{EQs:DipoleMoment}) we assumed the polar distortion
of the octahedral crystal field to be directed along the $z$ axis (i.e. with two oxygens on the symmetry axis),
which is not the case in MnWO$_4$, but this does not change our
semiquantitative model.

\section{Magnetic ordering in MnWO$_4$ with $\vec{k}=(1/4;1/2;1/2)$}

Equations~(\ref{EQs:Polarizationk=0}) give the electric
polarization for magnetic structures with $\vec{k}=0$. In MnWO$_4$
the magnetic phase transitions can be described by the order
parameters with $\vec{k}=(1/4;1/2;1/2)$~\cite{SakhnenkoMWO}. In
this point of the Brillouin zone the space group $P2/c$ possesses
two two-dimensional IR's $G_1$ and $G_2$. Using the magnetic
representation analysis conducted earlier~\cite{SakhnenkoMWO}, for
every direction $\alpha=x,y,z$ we introduce two order parameters
$(\eta_{1\alpha},\xi_{1\alpha})$ and
$(\eta_{2\alpha},\xi_{2\alpha})$ transforming according to $G_1$
and $G_2$, respectively. The spin components along $x$ induced by
these order parameters are given in table~\ref{tab:SpinStructures}.
\begin{table}
\caption{\label{tab:SpinStructures}
Components along $x$ of Mn$_1$ and Mn$_2$ spins induced by
$(\eta_{1x},\xi_{1x})$ and $(\eta_{2x},\xi_{2x})$ in the reference
unit cell $(0\cdot\vec{a}_1)$ and in the unit cell displaced by
one period of translation along $a$ ($1\cdot\vec{a}_1$). The last
two lines show the spin structures in each cell rewritten in terms
of $\vec{F}$ and $\vec{A}$.}
\begin{indented}
\item[]\begin{tabular}{@{}lll}
\br
& $0\cdot\vec{a}_1$& $1\cdot\vec{a}_1$\\
\mr
$S_{1x}$ & $\eta_{1x}+\eta_{2x}$ & $\xi_{1x}+\xi_{2x}$ \\
$S_{2x}$ & $-\xi_{1x}+\xi_{2x}$ & $\eta_{1x}-\eta_{2x}$\\
\mr
$F_x$ & $\eta_{1x}+\eta_{2x}-\xi_{1x}+\xi_{2x}$ & $\eta_{1x}-\eta_{2x}+\xi_{1x}+\xi_{2x}$ \\
$A_x$ & $\eta_{1x}+\eta_{2x}+\xi_{1x}-\xi_{2x}$ & $-\eta_{1x}+\eta_{2x}+\xi_{1x}+\xi_{2x}$ \\
\br
\end{tabular}
\end{indented}
\end{table}
For every unit cell one can rewrite the spin components in terms of
$\vec{F}$ and $\vec{A}$ as shown in the table. Analogous analysis
can be conducted for the $y$ and $z$ spin components.

The magnetic unit cell in MnWO$_4$ is 16 times the
crystallographic one. We now sum up $A_xF_x$ over the magnetic
cell $\frac{1}{16}\sum
A_xF_x=2(\eta_{1x}\eta_{2x}+\xi_{1x}\xi_{2x})$ to obtain the
electric polarization using~(\ref{EQs:Polarizationk=0})
\begin{equation}\label{EQs:Py_MWO}
P_{y}=d_{0y}\frac{2}{v}\left(\frac{\lambda}{\Delta}\right)^2(\eta_{1x}\eta_{2x}+\xi_{1x}\xi_{2x}).
\end{equation}
Equation~(\ref{EQs:Py_MWO}) gives $P_y$ in accordance with the
polarization that can be derived from the ME interaction
$P_y(\eta_{1x}\eta_{2x}+\xi_{1x}\xi_{2x})$ obtained from the
macroscopic symmetry analysis~\cite{SakhnenkoMWO}. In agreement
with the experiment~\cite{Taniguchi06} and phenomenological
model~\cite{SakhnenkoMWO} $P_y$ arises in the \textbf{AF2} phase
when both $G_1$ and $G_2$ condense. Long-wavelength modulation
does not lead to cancellation of~(\ref{EQs:Py_MWO}) as shown
earlier~\cite{SakhnenkoMWO}. Similar to~(\ref{EQs:Py_MWO}) other
contributions to polarization can be obtained using remaining ME
interactions~(\ref{EQs:MagnetoelectricInteractions}).

The numerical value of $P_y$ from~(\ref{EQs:Py_MWO}) for
MnWO$_4$ can be estimated as follows. We use the real crystal data
and oxygen positions from \cite{Lautenschlaeger} to
perform the crystal field expansion~(\ref{EQ:VcrystalField}) and
obtain $c_z\approx4.9\cdot10^{-9}$~N and $v\approx138$~\AA$^3$.
For the matrix elements $t_{\alpha,\gamma}$ we use the
hydrogen-like orbitals and obtain $t_{z,z^2}\approx0.67a_0/Z$,
where $a_0$ is the Bohr radius and $Z$ is the charge of the
nucleus and core electrons in units of $e$. Using $Z\approx5$,
$\lambda\approx0.05$~eV, $\Delta\approx1$~eV and noting that
$q_1\sim1$, $(\eta_{1x}\eta_{2x}+\xi_{1x}\xi_{2x})\sim1$ we obtain
$P_y\sim17$~$\mu$C/m$^2$ in good agreement with the experimental
value of the order of
$50$~$\mu$C/m$^2$~\cite{Taniguchi06,Arkenbout06}. In this
numerical estimate, however, we considered only one $e_g$ level
whereas other $3d$ electrons give comparable contributions to
$P_y$.

\section{Discussion}

Thus, we have built a microscopic model of ME interactions in
magnetoelectrics using MnWO$_4$ as an example. Starting with the
hypothetical magnetic order with $\vec{k}=0$ we determined the ME
interactions~(\ref{EQs:MagnetoelectricInteractions}). Noting that
Mn$^{2+}$ ions in wolframite are located in noncentrosymmetric
polar surroundings we suggested a microscopic model of
magnetoelectricity. The orbitals of $3d$ electrons of Mn$^{2+}$
ions possess electric dipole moments due to the crystal field
influence, which gives rise to additional contributions (such as
the considered $4p$ states) to their wave
functions~(\ref{EQ:PSIwavefunction}). The spin-orbit
coupling~(\ref{EQ:SOcoupling}) induces spin-dependent electric
dipole moments~(\ref{EQs:DipoleMoment}) since the angular momentum
operator $\vec{L}$ mixes different $|p_\alpha\rangle$ states. In
this part our approach combining phenomenological and microscopic
models resembles that suggested for the description of weak
ferromagnetism by Dzyaloshinskii~\cite{DzyaloshinskiiWeakFM} and
Moriya~\cite{Moriya}. Our approach differs from those of Sergienko et al.~\cite{SergienkoDMI}
and Katsura et al.~\cite{KatsuraSpinCurrent} who obtain the electric polarization as either $\vec{P}\sim\ [\vec{S}_i\times
\vec{S}_{i+1}]$ or $\vec{P}\sim\vec{e}_{ij} \times [\vec{S}_i\times\vec{S}_{j}]$, which is essentially a result of
interacting spins located at two different ions. On the contrary, in our model due to local noncentrosymmetry each magnetic ion
has spin-dependent electric dipole moment. We show that for certain spin configurations the sum over all magnetic ions of these spin-dependent electric dipole moments gives rise to macroscopic polarization.

We then apply this approach to describe real
magnetic structures in MnWO$_4$ with $\vec{k}=(1/4;1/2;1/2)$ by
summing up local contributions to polarization in every unit cell
of the magnetic cell. Our approach naturally accounts for the
macroscopic symmetry of the unit cell and is valid in both cases
of collinear (hypothetical) magnetic ordering with $\vec{k}=0$ and
complex long-wavelength modulated magnetic structure observed in
MnWO$_4$.

We have chosen MnWO$_4$ as an example since it has Mn$^{2+}$ ions
in polar surroundings and is directly applicable to our approach.
At the same time magnetoelectrics with magnetic ions in
noncentrosymmetric environment are numerous and a brief review of
the recently discovered multiferroics gives the following
examples. Similar to MnWO$_4$ the local C$_2$ symmetry is found
for Fe$^{3+}$ ions in NaFeSi$_2$O$_6$~\cite{JodlaukPyroxenes} and
for Ni$^{3+}$ ``spine'' spins in Ni$_3$V$_2$O$_8$~\cite{LawesNVO}.
Local C$_s$ symmetry is found for Cu$^{2+}$ ions in
LiCu$_2$O$_2$~\cite{ParkLiCu2O2} and for one of the Cr$^{3+}$
positions in $\alpha$-CaCr$_2$O$_4$~\cite{SinghCaCr2O4}. All of
the nonequivalent Fe$^{3+}$ ions in FeVO$_4$~\cite{DixitFeVO4} and
FeTe$_2$O$_5$Br~\cite{ZaharkoFeTe2O5Br} have local C$_1$ symmetry.
For all these magnetic ions the linear in $\vec{r}$ part of the
crystal field expansion can be written as
$V_{CF}(\vec{r})=V_0+\vec{c}\cdot\vec{r}$ with $|\vec{c}|$ taking
values from $9\cdot10^{-10}$~N for Ni$_3$V$_2$O$_8$ to
$6\cdot10^{-9}$~N for LiCu$_2$O$_2$.

Magnetoelectric effect was recently  found in manganese phosphorus
trisulfide MnPS$_3$~\cite{RessoucheMnPS3}. MnPS$_3$ possesses a
monoclinic crystal structure $C2/m$ and shows an antiferromagnetic
collinear order with $\vec{k}=0$ below T$_N$=78~K. Similar to
MnWO$_4$ it has two Mn$^{2+}$ ions in the unit cell with the local
C$_2$ symmetry and our analysis performed above for $\vec{k}=0$
magnetic structures in wolframite is directly applicable to
MnPS$_3$. According to the neutron diffraction data the magnetic
structure is characterized by $A_x$ and $A_z\neq0$ below T$_N$.
Thus, following the Eqs.~(\ref{EQs:MagnetoelectricInteractions})
one can expect linear magnetoelectric effect below T$_N$ with
magnetoelectric interactions $P_\mu A_xF_y$, $P_\mu A_zF_y$,
$P_yA_\mu F_\mu$, $P_yA_xF_z$ and $P_yA_zF_x$, where $\mu=x,z$.
According to Eqs.~(\ref{EQs:Polarizationk=0}) the magnetoelectric
coefficient $\alpha_{yz}=\rmd P_y/\rmd H_z$, for example, can be estimated
as $\alpha_{yz}=d_{0z}(1/v)(\lambda/\Delta)^2A_z\cdot
\rmd F_z/\rmd H_z$, with
$\rmd F_z/\rmd H_z\sim2\cdot10^{-6}$~Oe$^{-1}$~\cite{ToyoshimaMnPS3} and
$A_z\sim1$. The polar local distortion of the Mn$^{2+}$
environment in MnPS$_3$ is much smaller than in wolframite giving
$c_z\approx1.4\cdot10^{-11}$~N. Using $v\approx207$~\AA$^3$ we
obtain
$\alpha_{yz}\sim3.4\cdot10^{-8}$~$\mu$C$\cdot$m$^{-2}\cdot$Oe$^{-1}$,
which is rather small.

LiNiPO$_4$ possesses an orthorhombic symmetry with space group
$Pnma$ and shows linear magnetoelectric effect in the low
temperature $C$-type commensurate antiferromagnetic phase with
$\vec{k}=0$ below 20.8~K~\cite{KornevLiNiPO4,ToftPetersenLiNiPO4}.
The spins are predominately directed along the $c$ axis and the
magnetic structure is described by the order parameter $C_z$
transforming according to the IR GM$^{4-}$. The phenomenological
magnetoelectric interactions $C_zM_zP_x$ and $C_zM_xP_z$ were
suggested earlier~\cite{KornevLiNiPO4} and here we can estimate
the magnetoelectric coefficient according to our microscopic
model. Our microscopic approach differs from that suggested for
LiNiPO$_4$ earlier~\cite{JensenLiNiPO4}, which is based on
lowering the superexchange interaction energy due to the uniform
displacement of oxygen tetrahedra. The local symmetry of the
Ni$^{2+}$ ions surroundings is C$_s$ with
$|\vec{c}|\approx4.3\cdot10^{-9}$~N. Similar to the above case of
MnPS$_3$ using $\rmd M_x/\rmd H_x\sim2\cdot10^{-2}$~$\mu_{\rm B}$/T per
Ni-atom~\cite{ToftPetersenLiNiPO4} we obtain an estimation of the
value of magnetoelectric coefficient
$\alpha_{zx}\approx0.31$~$\mu$C$\cdot$m$^{-2}\cdot$T$^{-1}$ in
good agreement with the experimental value of
$0.2$~$\mu$C$\cdot$m$^{-2}\cdot$T$^{-1}$~\cite{KornevLiNiPO4} and
about three orders of magnitude higher than that in MnPS$_3$.

At the same time in many other magnetoelectrics magnetic ions are
located in centrosymmetric positions. Such is the case, for
example, in the rare-earth manganites RMnO$_3$~\cite{KimuraRMO3}
and CuO~\cite{KimuraCuO} where the local symmetry around Mn$^{3+}$
and Cu$^{2+}$ is C$_i$. Nevertheless, our approach is valid also
in these cases. The application of our model becomes more
complicated and will be published
elsewhere~\cite{SakhnenkoToBePublished}, but briefly can be
described as follows. The local symmetry around the rare-earth
ions in RMnO$_3$ is C$_s$. The importance of rare-earth ions in
formation of electric polarization in RMnO$_3$ was recently
pointed out~\cite{FeyerhermRMO3,SchierleRMO3}. Indeed, the
magnetic order of Mn$^{3+}$ ions induces magnetic ordering of the
rare-earths through various exchange mechanisms, which makes our
approach applicable. In the case of CuO the oxygens have local
symmetry C$_2$. The role of oxygen in the superexchange is still a
subject of debate~\cite{MoskvinPRB,MoskvinJETP}. Being the
intermediate ion conducting superexchange, O$^{2-}$ should possess
induced magnetic moment when copper spins order, which again
allows application of our model. The antiferromagnetic spin polarization at the oxygen sites
was measured, for example, in the multiferroic TbMn$_2$O$_5$~\cite{Beale_TbMn2O5}.

\section{Conclusions}

We have suggested a microscopic model of magnetoelectric interactions, which directly exploits the fact that in many magnetoelectrics magnetic ions are located in noncentrosymmetric positions. The model is illustrated by the examples of
MnWO$_4$ and LiNiPO$_4$, for which we obtained good correspondence of the values of electric polarization and magnetoelectric coefficient, respectively. We also give an estimate of the magnetoelectric coefficient in MnPS$_3$.


\section*{References}

\begin{thebibliography}{10}

\bibitem{Curie}
Curie P.
\newblock {\em J. Physique}, 3:393, 1894.

\bibitem{Dzyaloshinskii}
Dzyaloshinskii~I E.
\newblock {\em Sov. Phys.-JETP}, 10:628, 1959.

\bibitem{Astrov}
Astrov~D N.
\newblock {\em Sov. Phys.-JETP}, 11:708, 1960.

\bibitem{ReviewSpiralTokura}
Tokura Y and Seki S.
\newblock {\em Adv. Mater.}, 22:1554, 2010.

\bibitem{ArimaSpinDrivenFerro}
Arima T.
\newblock {\em J. Phys. Soc. Jap.}, 80:052001, 2011.

\bibitem{SakhnenkoImproperFerroelectric}
Sakhnenko~V P and Ter-Oganessian~N V.
\newblock {\em Ferroelectrics}, 400:12, 2010.

\bibitem{Kimura125manganites}
Kimura H, Kobayashi S, Wakimoto S, Noda Y, and Kohn K.
\newblock {\em Ferroelectrics}, 354:77, 2007.

\bibitem{SergienkoDMI}
Sergienko~I A and Dagotto E.
\newblock {\em Phys. Rev. B}, 73:094434, 2006.

\bibitem{KatsuraSpinCurrent}
Katsura H, Nagaosa N, and Balatsky~A V.
\newblock {\em Phys. Rev. Lett.}, 95:057205, 2005.

\bibitem{MochizukiSpinCurrent}
Mochizuki M and Furukawa N.
\newblock {\em Phys. Rev. Lett.}, 105:187601, 2010.

\bibitem{MoskvinPRB}
Moskvin~A S and Drechsler S-L.
\newblock {\em Phys. Rev. B}, 78:024102, 2008.

\bibitem{MoskvinEPJB}
Moskvin~A S and Drechsler S-L.
\newblock {\em Eur. Phys. J. B}, 71:331, 2009.

\bibitem{JiaOnodaNagaosaMOMcluster}
Jia C, Onoda S, Nagaosa N, and Han~J H.
\newblock {\em Phys. Rev. B}, 74:224444, 2006.

\bibitem{Lautenschlaeger}
Lautenschl{\"{a}}ger G, Weitzel H, Vogt T, Hock R, Bohm A, Bonnet M, and Fuess
  H.
\newblock {\em Phys. Rev. B}, 48:6087, 1993.

\bibitem{Taniguchi06}
Taniguchi K, Abe N, Takenobu T, Iwasa Y, and Arima T.
\newblock {\em Phys.\ Rev. Lett.}, 97:097203, 2006.

\bibitem{SakhnenkoMWO}
Sakhnenko~V P and Ter-Oganessian~N V.
\newblock {\em J. Phys.: Condens. Matter}, 22:226002, 2010.

\bibitem{Arkenbout06}
Arkenbout~A H, Palstra T~T M, Siegrist T, and Kimura T.
\newblock {\em Phys. Rev. B}, 74:184431, 2006.

\bibitem{DzyaloshinskiiWeakFM}
Dzyaloshinskii I.
\newblock {\em J. Phys. Chem. Solids}, 4:241, 1958.

\bibitem{Moriya}
Moriya T.
\newblock {\em Phys. Rev.}, 120:91, 1960.

\bibitem{JodlaukPyroxenes}
Jodlauk S, Becker P, Mydosh~J A, Khomskii~D I, Lorenz T, Streltsov~S V, Hezel~D
  C, and Bohat{\'{y}} L.
\newblock {\em J. Phys.: Condens. Matter}, 19:432201, 2007.

\bibitem{LawesNVO}
Lawes G, Harris~A B, Kimura T, Rogado N, Cava~R J, Aharony A, Entin-Wohlman O,
  Yildirim T, Kenzelmann M, Broholm C, and Ramirez~A P.
\newblock {\em Phys.\ Rev. Lett.}, 95:087205, 2005.

\bibitem{ParkLiCu2O2}
Park S, Choi~Y J, Zhang~C L, and Cheong S-W.
\newblock {\em Phys.\ Rev. Lett.}, 98:057601, 2007.

\bibitem{SinghCaCr2O4}
Singh K, Simon C, and Toledano P.
\newblock {\em Phys. Rev. B}, 84:064129, 2011.

\bibitem{DixitFeVO4}
Dixit A and Lawes G.
\newblock {\em J. Phys.: Condens. Matter}, 21:456003, 2009.

\bibitem{ZaharkoFeTe2O5Br}
Zaharko O, Pregelj M, Ar{\v{c}}on D, Brown~P J, Chernyshov D, Stuhr U, and
  Berger H.
\newblock {\em J. Phys.: Conf. Ser.}, 211:012002, 2010.

\bibitem{RessoucheMnPS3}
Ressouche E, Loire M, Simonet V, Ballou R, Stunault A, and Wildes A.
\newblock {\em Phys. Rev. B}, 82:100408(R), 2010.

\bibitem{ToyoshimaMnPS3}
Toyoshima W, Masubuchi T, Watanabe T, Takase K, Matsubayashi K, Uwatoko Y, and
  Takano Y.
\newblock {\em J. Phys.: Conf. Ser.}, 150:042215, 2009.

\bibitem{KornevLiNiPO4}
Kornev I, Bichurin M, Rivera J-P, Gentil S, Schmid H, Jansen A~G M, and Wyder
  P.
\newblock {\em Phys. Rev. B}, 62:12247, 2000.

\bibitem{ToftPetersenLiNiPO4}
Toft-Petersen R, Jensen J, Jensen T~B S, Andersen~N H, Christensen~N B,
  Niedermayer C, Kenzelmann M, Skoulatos M, Le~M D, Lefmann K, Hansen~S R,
  Li~J, Zarestky~J L, and Vaknin D.
\newblock {\em Phys. Rev. B}, 84:054408, 2011.

\bibitem{JensenLiNiPO4}
Jensen T~B S, Christensen~N B, Kenzelmann M, R{\o}nnow~H M, Niedermayer C,
  Andersen~N H, Lefmann K, Schefer J, Zimmermann~M v, Li~J, Zarestky~J L, and
  Vaknin D.
\newblock {\em Phys. Rev. B}, 79:092412, 2009.

\bibitem{KimuraRMO3}
Kimura T, Lawes G, Goto T, Tokura Y, and Ramirez~A P.
\newblock {\em Phys. Rev. B}, 71:224425, 2005.

\bibitem{KimuraCuO}
Kimura T, Sekio Y, Nakamura H, Siegrist T, and Ramirez~A P.
\newblock {\em Nat. Mater.}, 7:291, 2008.

\bibitem{SakhnenkoToBePublished}
Sakhnenko~V P and Ter-Oganessian~N V.
\newblock To be published.

\bibitem{FeyerhermRMO3}
Feyerherm R, Dudzik E, Prokhnenko O, and Argyriou~D N.
\newblock {\em J. Phys.: Conf. Ser.}, 200:012032, 2010.

\bibitem{SchierleRMO3}
Schierle E, Soltwisch V, Schmitz D, Feyerherm R, Maljuk A, Yokaichiya F,
  Argyriou~D N, and Weschke E.
\newblock {\em Phys.\ Rev. Lett.}, 105:167207, 2010.

\bibitem{MoskvinJETP}
Moskvin~A S.
\newblock {\em JETP}, 104:913, 2007.

\bibitem{Beale_TbMn2O5}
Beale T~A W, Wilkins~S B, Johnson~R D, Bland~S R, Joly Y, Forrest~T R,
  McMorrow~D F, Yakhou F, Prabhakaran D, Boothroyd~A T, and Hatton~P D.
\newblock {\em Phys. Rev. Lett.}, 105:087203, 2010.

\end{thebibliography}


\end{document}